**Sharper Than Ever:**
**Do Modern Observations Pin Down the Solar Radius to Converge on New Standards?**

Jean-Pierre Rozelot[1], Alexander Kosovichev[2,3]

[1] Université de la Côte d'Azur, Grasse 06130, France
email: jp.rozelot@orange.fr

[2] Center for Computational Heliophysics, New Jersey Institute of Technology,
University Heights, Newark, NJ 07102, U.S.A.
[3] NASA Ames Research Center, Moffett Field, CA 94035 U.S.A.
email: alexander.g.kosovichev@njit.edu

**Abstract :**

Solar radius measurements and their variations - if any - are a difficult problem that has vexed researchers for decades. In this paper, we have attempted to clarify the various ways of expressing the definition "solar diameter", from a physical point of view. The concept of diameter is taken here in its broadest sense, leaving aside the issue concerning the oblateness caused by surface and internal angular velocity variations, as deviations from sphericity are negligible in our context. Astrometric time-series observations are still needed, and we advocate strengthening long-term metrological measures to achieve greater consensus on the subject. To date, modern observations of the solar diameter provide a frame of reference, and we give a new glossary. By comparing the best values obtained to date, it is shown that the "seismic radius" obtained from the Solar and Heliospheric Observatory (SOHO) and the Solar Dynamics Observatory (SDO) provides the best determination, a finding supported by observations made at the Calern (F) and Pic du Midi (F) observatories. The latest results on the leptocline show that it is more important than ever to consider at which layers of the Sun radius measurements are carried out. On this basis, we hope to converge on new standards.

**Introduction.**

The approximately five decades from the 1960s to the 2010s saw intense scientific activity focused on achieving the most accurate possible measurements of the solar diameter. This relative excitement, which focused as much on observations as on theoretical works, centered mainly on three main issues. (i) The determination of the solar diameter from helioseismic data in the 1990s, which led to a discrepancy with the so-called photospheric radius, has also been the focus of numerous measures. (ii) The solar diameter variation over time, in relation to the activity cycle and its possible implication in irradiance variability. A third point concerns the variability of the solar diameter with heliographic latitude, from a static or dynamical point of view, which leads not only to the flattening but also to the concept of gravitational moments and their relativistic effects. Underlying these three issues is a need for a physically based definition of a solar diameter, which, while not really debatable, remains subject to some caveat. Such matters have been widely discussed in numerous articles. Indeed, at the turning point of the 1970s, there was even an overabundance, which likely did more harm than good to the results. To date, is it possible to reach a consensus on this subject?

Hereafter, we refer to the solar radius instead of its diameter. It corresponds to what is commonly used today but historically stems from observations made with the solar astrolabe. Due to the design of this instrument, measurements do not give a unique diameter but two radii forming a small angle epsilon between them (see Rozelot and Damiani, 2012, p. 29). This misuse of language persists today, considering the Sun as a whole, but is comprehensible in the scope of the flattening (or changing shape with latitude, see for ex. Jain, 2025 or Rozelot et al., 2020).

The first accurate measurements of the solar radius were made by Mouton in 1610, Picard (XVIIth century), and Secchi (at the end of the XVIIIth century). Their measurements remain relevant today, as it has been pointed out that the size of the Sun may have varied over time. For example, Ribes et al. (1987) showed that the solar diameter was larger and the solar rotation slower during the Maunder Minimum (MM) period, an issue revisited by Zhang et al. (2015), who highlighted the importance of conducting long-term measurements.

This paper is divided into two sections. In the first one, we recall the different astrometric determinations of the solar radius, focusing on the methods and current limits. The second one is devoted to generic types of solar radius; a synthetic table is proposed as a conclusion.

## 1. Solar radius determination, methods, and current limits

The solar radius $R_\odot$ is a fundamental astrophysical parameter entering structure models, and acts as a reference in astrophysical measurements. A change in its absolute value induces a change in quantities referred to by it, for example, the size of stars; moreover, it may have implications in the stellar stratification. However, determining the solar radius with high accuracy is nontrivial. Because of the presence of flows and oscillations on small scales, the surface of the Sun has no sharp physical edge. The height of the solar limb thus depends on the dynamics of the upper solar atmosphere at a given wavelength (e.g., Rozelot et al. 2015; Battaglia et al., 2017).

Strictly speaking, it should be noted that the value of $R_\odot$, taken as a standard, refers to the equatorial radius of the star $R_{eq}$, since the latter is fluid and deforms under the action of rotation. The resulting flattening is small (about 7.81 mas, theoretically), but it does exist and has been measured (about 8.2 mas). In addition, the subsurface layers are not homogeneous and are impregnated with sometimes intense magnetism, leading to various effects, such as shearing and zonal flows (torsional oscillations). As a result, the measured "radius" value varies with latitude (theta), being different from equatorial and polar values (e.g. Kosovichev et al, 1998; Rozelot and Damiani, 2011). Consequently, it would be logical to specify the time being referred to, since all these "radii" may vary over time, even if these changes are rather faint. It is also important to emphasize the wavelength (lambda) used, since the diameter of a celestial body is not the same in white light as it is at a given wavelength. Finally, the values deduced from observations have been calculated using reference values, such as the astronomical unit, which have evolved over time as celestial parameters have become more accurate. The reference system used should therefore be indicated. Although all the resulting differences could be judged minimal, from a metrological accuracy perspective, they must be highlighted.

Regardless of the method used, the solar radius measured should be the same, apart from experimental errors. We know that this is not the case, which raises interesting questions related to the internal structure of the near-surface layers. Based on radiative transfer calculations, an attempt to reconcile the discrepancy between the seismic radius and its conventional value deduced from observations of the intensity profile at the limb has been made by Haberreiter et al., 2008. However, this approach does not explain why other measurements carried out by other means lead to different results, except that different atmospheric models would have to be used (without ruling out the influence of subsurface magnetism, which has so far been given very limited consideration, see eg. Bétrisey et al., 2025). Thus, the discrepancies between the different types of solar radius measurements cannot yet be explained by existing solar models. For all these reasons, we felt it was important to take stock of this article devoted to solar radius measurements. A summary history has been provided by Rozelot and Damiani (2012); Table 1 contains some errors that are corrected in a specific note (Rozelot et al., 2026).

## 2. Generic types of solar radius. An attempt to set clear standards

### 2.1 The photospheric radius.

Two methods are usually employed for measuring the solar radius. The first one, known as transit time passage (generally at noon, in front of a fiduciary line used to measure the time between two east-west contacts), is carried out by means, in the past, of meridian circles, micrometers, and, more recently, by solar astrolabes (visual, then CDD). Linked to this method are determinations during solar eclipses, and the transits of planets in front of the Sun (Mercury, Venus). The second method refers to angular measurements using heliometers, spectroscopy tools, or dedicated space instruments (e.g., SDS, SOHO, RHESSI, SDO, PICARD). They lead to the determination of the so-called photospheric radius $R_{ph}$.

The transit time method used during the transits of Venus across the Sun in 1874 and 1882 enabled Auwers (1891) to provide a reliable value for the diameter of the Sun, which was adopted as the reference value for 120 years (1895-2015), also known as the "canonical" value:

$[R_\odot]$ *canonique (1895-2015)* = 965.63 ± 0.05 arcsec.

The astronomical unit in force in 1891 had a value of 149 597 900 km. This results in a solar radius of 695 996 (914) km. However, at the 2012 IAU General Assembly in Beijing, the Astronomical Unit was revised to 149 597 870.70 km, a difference of 29.30 km from the previous value. This gives a radius of 695 996 (773) km, counting as a difference of 141 m. This may seem insignificant, given the measurement uncertainties, but it is nonetheless real. The value adopted by consensus until 2015 was:

$[R_\odot]$ *canonique (1895-2015)* = 695 997 ± 36 km

Based on measurements taken using the Solar Diameter Monitor (SDM) at a wavelength of 800 nm (± 10) over the period 1981-1987, Brown and Christensen-Dalsgaard (1998) obtained a value for the photospheric radius of

$R_{ph}$ = 695 508 ± 26 km or $R_{ph}$ = 958.96 ± 0.036 arcsec,

taking into account the old AU value (which is logical, given the dates on which the results were obtained). This latter value has been widely used, as it was the first time that the introduction of automated measurements into the observation system eliminated the possibility of human observation bias. This results in a difference of 489 km (less) from Auwers' value. It is this value $R_{ph}$ that is found in Allen's "Astrophysical Quantities" (from the fourth edition in 2000 to the last one revised in 2013).

### 2.2. The seismic radius:

The seismic radius is the effective solar radius inferred from helioseismic measurements (GONG ground-based data or space data).

#### 2.2.1. The fundamental seismic radius (f-modes).

Schou et al. (1997) reported on the first helioseismic determination of the solar radius using high-precision measurements of oscillation frequencies of the fundamental (f) mode of the Sun, obtained from the Michelson Doppler Imager (MDI) experiment on board the *SOHO* spacecraft (Scherrer et al. 1996). They found in the medium angular degree range, l = 88–250:

$[R_{sis\,f}]$ = 695 680 ± 30 km.

This represents a difference of 317 km (less) compared to Auwers' value and 172 km (greater) compared to Brown's value. It turns out that

$[R_{sis\,f}]$ = 959.20 ± 0.04 arcsec,

(taking into account the new value of the UA, but 959.19 arcsec with the old one, a difference of about 136 m, as seen before).

In the same way, Tripathy and Antia and (1999), using data extracted from the GONG network, estimated the solar radius to be

$[R_{sis\,f}]$ = 695 770 ± 100 km or 959.32 ± 0.14 arcsec.

This amounts to a difference of 90 km (greater), from the work of Schou, 227 km (less) from that of Auwers, and 262 km (greater) from that of Brown. It was noted at that time that the exact extent of the reduction in radius (from Auwers to Schou) to match the f-mode frequencies results will depend on the input physics used in constructing the solar model. In this way (and as seen before), Habereiter et al. (2008) were able to reconcile the values to some extent using a radiative transfer code combined with a solar atmosphere code in spherical symmetry. Still, this approach needs to be reviewed alongside other models, as such analysis depends on atmospheric models.

### 2.2.2. The acoustic seismic radius (p-modes)

As early as 2001, Takada and Gough (2001) indicated that the solar radius itself can be constrained by the p-mode frequencies. Using SOHO/MDI data, they estimated the radius as 695 690 ± 140 km. Considering that p modes react differently from f modes to the various processes in the upper convective boundary, Takada and Gough (2024) showed that the acoustic determination is robust. Similar to Schou et al. (1997), the radius is calibrated by the distance from the centre of the Sun to the position in the subphotospheric layers where the first derivative of the density scale height changes essentially discontinuously. On this basis, and assuming a homologous difference in the structure of the outer layers of the Sun beyond the upper turning point of the p modes, they obtained

$[R_{sis\ a}]$ = 695 780 ± 160 km or 959.33 ± 0.22 arcsec,

according to the "old" value of the UA. It turns out that this estimate is 217 km lower than Auwers's determination, 272 km greater than Brown's, 100 km lower than Schou's, and of the order of Antia's result (10 km less, within the errors).

### 2.3. Determination of the solar radius from space

One might think that direct spatial measurements (i.e., through limb-shape accurate profile), by eliminating the effects of the Earth's atmosphere, would provide very accurate values for the solar radius. Several experiments have been carried out aboard balloons (SDS by Sofia Sabatino (2013) as early as 1991) or satellites (SOHO, SODISM, RHESSI, SDO, etc.), with success if we consider only the results obtained, but less so if we look at their intrinsic properties in terms of consistency. The difficulties arise from optical distortions, annual temperature change due to the variable Sun-satellite distance, secular variation due to degradation of the front window transparency, filter degradation with time, etc.

One of the first precise solar astrometry determinations was made with the help of the Michelson Doppler Imager (MDI, on board the Solar and Heliospheric Observatory (SOHO), which enabled Emilio et al. (2010) to measure, at the wavelength of 677 nm, the absolute value of the solar radius as:

$R_{spa}$= 959.28 ± 0.15 arsec i.e. 695 743 ± 109 km

This is greater than the ground-based measurements made by Brown and Christensen-Dalsgaard (1998), of 235 km and 37 km less than the acoustic radius.

A special mention should be made of RHESSI (The Reuven Ramaty High-Energy Solar Spectroscopic Imager), which enabled Battaglia et al. (2017) to measure the solar radius as $R_{X-ray}$ = 964.05 ± 0.15 arcsec (699 202 ± 109 km) in the energy range 0.05–0.4 nm. This provides an interesting point for calibrating diameter measurements above or below the reference limb.

Numerous other measurements by different means and at different dates which can be found in a range of papers and reviews (a list is provided, for example, in Rozelot et al. (2015); work is underway to untangle all these values, classifying them, for example, by method of acquisition (ground, space, eclipse), by measurement category (transit time, astrolabes, heliometer...), methods of helioseismology, and others).

### 2.4. Ground-based determination of the solar radius

As previously seen, considerable efforts have been made in the latter part of the 21st century, but given the precision required, instruments were still at the cutting edge of the techniques. Moreover, ground-based observations suffer from degradation of the solar signal by atmospheric turbulence, and require numerous corrections (instrumental distortions, refraction, scintillation, blurring effects…). A description of the various methods used, and a critical overview can be found in Rozelot and Damiani (2012).

Let us mention only, as examples (i) the CDD solar astrolabe results (Calern observatory -F) from Morand et al. (2011), who found, at a wavelength of 540.0 nm:

959.48 ± 0.01 arcsec, i.e . 695 888 ± 7.25 km

over the 1985 - 2009 time period, one of the longest-running series so far achieved; (ii) those of Rozelot et al. (2003) obtained at the Pic du Midi observatory (F), at a wavelength of 505.8 nm:

959.434 ± 0.008 arcsec, i.e. 695 855 ± 5.8 km.

These last measurements (fast photoelectric scans of the opposite limbs of the Sun recorded quasi-simultaneously to freeze atmospheric fluctuations) were obtained during four days of exceptional meteorological conditions (a mean seeing of 18 cm from September 3 to 6, 2001), during which the diffraction limit of the refractor was reached. The two values (Calern and Pic du Midi) differ by 0.046 arcsec, or 33 km, and the Pic du Midi value differs (less) from Takada and Gough's (2015) result by 75 km only; therefore, it is included in the 160 km error as determined by these authors.

### 2.5. Eclipse astrometry revival (2024) as a means for radius measurements

Modern solar eclipses have once again become powerful tools, thanks to modern high-resolution lunar limb profiles from Kaguya and LRO/LOLA. These datasets revolutionized eclipse-based radius measurements, reducing uncertainties by orders of magnitude and enabling sub-0.1-arcsecond determinations. They offer a distinct advantage: extremely sharp limb definition, which can be determined at a specific wavelength and in a very narrow bandwidth. But also, some limitations: rare events, dependence on lunar topography accuracy. Recent results (2010-2015) led to a mean solar radius of (Lamy et al, 2015):

$R_{ecl}$= 696 246 ± 45 km or 959.97 ± 0.06 arcsec.

Of significance in this regard is the latest historical determination of the solar radius during the eclipse of 1715 by Hayakawa et al. (2026), which is 696 250 ± 170 km (959.98" ± 0.23"), lending such measurements importance for long-term studies (as seen also in the introduction).

## 2.6. The nominal radius

Relying mainly on the helioseismic definition of the solar radius and measurements of Schou et al. (1997), but also other rounded off measurements, the International Astronomical Union General Assembly in Honolulu (USA), recommended in 2015 that the solar radius $R_N$ (Nominal) must be (Prša et al. 2016 ):

$\mathcal{R}_{\odot}^{\mathrm{N}}$ *Nominal* = 6.957 $10^8$ m or 959.22 arcsec ( 959.23, confounding the angle and its tangent)

This accounts for a difference of 297 km or 0.41 arcsecond compared with the Auwers value (lower), 192 km with that of Brown (greater), 20 km with that of Schou (lower), and 70 km with Antia (greater).

This “nominal” value has the advantage of being supported by the IAU. It does not require indicating the depth at which the solar radius must be referred, nor the reference date, nor its heliographic latitude. From the point of view of stellar structure, it provides a “nominal” standard used to calibrate the standard solar and stellar models (e.g., Christensen-Dalsgaard, 2021), with such specifications, for instance, to characterize the spatial distributions of chemical composition, temperature, and density.

In this respect, we emphasize the physical conditions prevailing in the Sun’s subsurface layers that are of crucial importance. These conditions are of interest primarily for helio- and asteroseismology, because subsurface layer dynamics affect properties of acoustic modes (the so-called “surface effects”), making it difficult to match observed and modeled frequencies. This is a key aspect not only for the Sun but also for any solar-type oscillators (where related issues with parameter constraints can worsen than in the solar case, as surface effect corrections are usually calibrated with the Sun as a reference, see e.g. Bétrisey et al. 2025).

Table 1. Proposed new glossary of the various solar radii in use.
Adapted from Takada & Gough (2024).

| **Radius name** | **Label** | **Meaning** |
|---|---|---|
| Photospheric | $R_{ph}$ | **a/** Distance from the solar center to the photospheric surface, defined as the layer where the optical depth $\tau$ = 1 for a particular wavelength, usually 500 nm.<br><br>**b/** Defined by the depth where the temperature equals the effective temperature:<br>$R_{\mathrm{ph}}^2 T_{\mathrm{eff}}^4 = L_{\odot}/(4\pi\sigma)$.<br><br>**c/** Distance measured at the inflection point, for a given wavelength.<br><br>• $R_{ph}$ must be focused on the wavelength dependence, i.e., the height or depth at which the measurement is made. |

| Canonical | $R_c$ | Adopted radius through a consensus (1895-2015). |
|---|---|---|
| Nominal | $\mathcal{R}_{\odot}^{N}$ | Adopted radius by the IAU GA (2015), chosen to standardize calculations, not to match each new best measurement. |
| Generic | $R_{\odot}$ | **a/** Generic term, without reference to any particular concept of time, depth, latitude... <br><br> **b/** Used to calibrate models. |
| Seismic: | $R_s$ | Radius calculated using helioseismology, f_modes: fundamental -- p_modes: acoustic |
| 1. Fundamental | $R_f$ | Fundamental photospheric radius scaled through f-modes, which is the distance from the solar center to the center of energy of each f-mode (essentially the peak in the kinetic-energy distribution). |
| 2. Acoustic | $R_{ac}$ | Acoustic photospheric radius scaled through p-modes, which is the distance from the center to the subphotospheric layer where the acoustic cut-off frequency changes extremely rapidly. |

**Conclusion**

A reliable measurement of the solar diameter does not yet seem to have reached a consensus. Here we have attempted to distinguish among the different designations of the radius, giving them physical meanings, summarized in Table 1.

As regards the nominal value, recognized by the IAU in 2015, unlike the photospheric measurements, this definition is not based on optical parameters such as wavelength, temperature, etc., nor on geometric parameters (latitude, etc.) or date criteria. It is derived from helioseismological measurements and serves as a standard for calibrating solar and stellar models or for other applications concerning the solar and stellar structure and evolution.

It is worth noting the agreement between the values determined on the ground (Calern; Pic du Midi), at a wavelength of 520 ± 20 nm, and those obtained from acoustic modes by Takada and Gough (2024), as the difference is only some 75 km, inside the current errors. Such a value is one of the best determinations of the solar radius to date.

By applying a helioseismic inversion technique to the observed variations of f-mode frequencies, it has been shown that the seismic radius changes are associated with variations in the subsurface stratification (Rozelot et al., 2026). The strongest variations appear just below the surface, around 0.995 $R_{\odot}$ (about 3.5 Mm below the surface), in antiphase with solar activity, changing in phase in the deeper layers of the Sun (between 0.975 and 0.99 $R_{\odot}$). Such variations in the leptocline stratification can be caused by subsurface magnetic fields and changes in the temperature distribution. This may explain differences in results obtained by different means of observation, when not probing the same atmospheric layers, which have sometimes been the source of controversy in the past.

To be complete, relativistic corrections would have to be applied, which would reduce the absolute value of the solar diameter by 1.5 km (Castellani et al. 1999).

These questions are not trivial, and it is reasonable to assume that new measurements of the solar diameter, called for by the authors, for example, at DKIST in Hawaii (USA), implementing ultra-high-resolution limb profiles and refined wavelength-dependent models, will unlock this centuries-old mystery.

**Acknowledgments**

The authors thank the referee for his/her relevant comments.

The work was partially supported by NASA grants: 80NSSC20K1320 and 80NSSC22M0162.